# SUB-CLUSTERING AMONG LOCAL GROUP GALAXIES

SIDNEY VAN DEN BERGH

Dominion Astrophysical Observatory
Herzberg Institute of Astrophysics
National Research Council
5071 West Saanich Road
Victoria, British Columbia, V8X 4M6, Canada
vdB@dao.nrc.ca

## ABSTRACT

A new compilation of data on Local Group members shows that low-luminosity early-type galaxies are, at the 99% confidence level, more strongly concentrated in the M 31, and Galactic sub-clumps, than are dwarf irregular galaxies. It is not yet clear if star formation continued longer in dSph galaxies that are located far from M 31 and the Galaxy, than it did in those that are not associated with the Andromeda and Milky Way sub-clumps of the Local Group.

1.   INTRODUCTION

It has been known for many years (van den Bergh 1968), that most of the luminous members of the Local Group (Hubble 1936) are concentrated in two sub-clusters. One of these is centered on M 31, and also contains M 32, M 33, NGC 147, NGC 185 and NGC 205. The other sub-clustering is centered on the Milky Way system, and also encompasses the two Magellanic Clouds. The galaxies NGC 6822 and IC 1613 were found not to be associated with the Galactic and M 31 sub-clusters of the Local Group. During recent years many new fainter galaxies have been found to be associated with the Local Group. It is the purpose of the present note to examine the sub-clustering of these fainter Local Group members. Furthermore an attempt is made to study the dependence of this sub-clustering on the morphological types of these faint galaxies.

2.   SUB-CLUSTERING AMONG FAINT GALAXIES

The data on which this investigation is based are given in Table 1, which is taken from information listed in the monograph The Galaxies of the Local Group (van den Bergh 2000). Some of this material has already been discussed in van den Bergh (1999). Listed in Table 1 are all of the Local Group galaxies fainter than $M_V = -16.5$, i.e. those that are less luminous than the elliptical galaxy M 32. Two of the galaxies in Table 1 (Aquarius and SagDIG) are so distant that they might turn out not to be physical members of the Local Group. Furthermore, Leo I has such a large radial velocity, and is so distant from the Galaxy, that it may not be a satellite of the Milky Way System. In Tables 2 and 3 it has therefore been listed under "Gal ?". Finally, it is not certain that IC 10 and Pisces are members of the M 31 sub-group of the Local Group, so they are listed as "M 31 ?". In Tables 2 and 3 Aquarius, which is too faint to have a reliable morphological stage classification on the DDO system, has been counted under "dIr/dSph". Table 2 shows the relation between the morphological types of the galaxies listed in Table 1, and their assignment to sub-groups. The table shows that dSph galaxies are strongly concentrated in the M 31 and Galaxy sub-groups, whereas the dIr galaxies are mostly not associated with these two sub-groups. A $\chi^2$ test shows that the relation between morphological type and environment is significant at $>99.9\%$. It should, however, be emphasized that the degree of clustering by very faint galaxies may have been overestimated. This is so because (1) dim resolved galaxies, like Dra and UMi, can probably only be discovered close to the Galaxy, and (2) low surface brightness galaxies, such as And I - And VII, were discovered during intensive searches in the neighborhood of M 31. This bias may be avoided (or greatly reduced) by only considering those dwarf galaxies that are more luminous than $M_V = -11.0$. Table 3 shows the relation between morphological type and environment for the Local Group galaxies with luminosities in the range $-16.5 < M_V < -11.5$. The data in this table again exhibit a strong correlation between morphological type and environment. A $\chi^2$ test shows this correlation to be significant at the 99% level. It is therefore concluded that presently available data show that low-luminosity early-type Local Group galaxies are much more strongly sub-clustered, than are low-luminosity late-type galaxies. Such an effect was first noted by Einasto et al. (1974). These authors attributed this effect to ram-pressure stripping of interstellar material by the gaseous coronae of massive galaxies. However, we now know that this is not likely to be the whole story, because many dIr galaxies rotate, whereas the majority of dSph galaxies do not. It would be interesting to know if dSph galaxies, such as Dra and UMi, in which most star formation took place long ago (Mateo 1998), are more tightly concentrated around their parent galaxies, than objects such as Carina, in which the rate of star formation peaked only ~ 6 Gyr ago.

# TABLE 1

## SUB-CLUSTER ASSIGNMENTS FOR FAINT LOCAL GROUP GALAXIES

| Name | Alias | Type | $(m-M)_o$ | $M_V$ | Sub-cluster |
|---|---|---|---|---|---|
| NGC 205 | ... | Sph | 24.4 | -16.4 | M 31 |
| IC 10 | ... | Ir IV: | 24.55 | -16.3 | M 31 ? |
| NGC 6822 | ... | Ir IV-V | 23.5 | -16.0 | x |
| NGC 185 | ... | Sph | 24.1 | -15.6 | M 31 |
| IC 1613 | ... | Ir V | 23.3 | -15.3 | x |
| NGC 147 | ... | Sph | 24.1 | -15.1 | M 31 |
| WLM | DDO 221 | Ir IV-V | 24.85 | -14.4 | x |
| Sagittarius | | dSph(t) | 17.0 | -13.8:: | Gal |
| Fornax | ... | dSph | 20.7 | -13.1 | Gal |
| Pegasus | DDO 216 | Ir V | 24.4 | -12.3 | x |
| Leo I | Regulus | dSph | 22.0 | -11.9 | Gal ? |
| And I | ... | dSph | 24.55 | -11.8 | M 31 |
| And II | ... | dSph | 23.8 | -11.8 | M 31 |
| Leo A | DDO 69 | Ir V | 24.2 | -11.5 | x |
| Aquarius[a] | DDO 210 | V | 25.05 | -11.3 | x |
| SagDIG[a] | ... | Ir V | 25.7: | -10.7: | x |
| Pegasus II | And VI | dSph | 24.45 | -10.7 | M 31 |
| Pisces | LGS 3 | dIr/dSph | 24.55 | -10.4 | M 31 ? |
| And V | ... | dSph | 24.55 | -10.2 | M 31 |
| And III | ... | dSph | 24.4 | -10.2 | M 31 |
| Leo II | ... | dSph | 21.6 | -10.1 | Gal |
| Phoenix | ... | dIr/dSph | 23.0 | -9.8 | x |
| Sculptor | ... | dSph | 19.7 | -9.8 | Gal |
| Cassiopeia | And VII | dSph | 24.25 | -9.6 | M 31 |
| Tucana | ... | dSph | 24.7 | -9.6 | x |
| Sextans | ... | dSph | 19.7 | -9.5 | Gal |
| Carina | ... | dSph | 20.0 | -9.4 | Gal |
| Draco | ... | dSph | 19.5 | -8.6 | Gal |
| Ursa Minor | ... | dSph | 19.0 | -8.5 | Gal |

[a]   Possible Local Group member

## TABLE 2

## CLASSIFICATION TYPE VERSUS ENVIRONMENT[a]

| Environment | Sph + dSph | dSph/dIr | Ir |
|---|---|---|---|
| Isolated | 1 | 2 | 6 |
| ? | 1 | 1 | 1 |
| in sub-cluster | 15 | 0 | 0 |

[a] All Local Group galaxies with $M_V > -18.5$.

## TABLE 3

## CLASSIFICATION TYPE VERSUS ENVIRONMENT[a]

| Environment | Sph + dSph | dSp/Ir + ? | Ir |
|---|---|---|---|
| Isolated | 0 | 1 | 5 |
| ? | 1 | 0 | 1 |
| In sub-cluster | 7 | 0 | 0 |

[a]    Local Group galaxies with $-16.5 < M_V < -11.0$.

- 8 -## REFERENCES

Einasto, J., Saar, E., Kaasik, A., & Chernin, A.D. 1974, Nature, 252,111

Hubble, E. 1936, The Realm of the Nebulae, (New Haven: Yale University Press)

Mateo, M. 1998, ARAA, 36, 435

van den Bergh, S. 1968, Comm. David Dunlap Obs. No. 195

van den Bergh, S. 1999, in Stellar Content of the Local Group, (= IAU Symposium No. 192), Eds. P. Whitelock and R.D. Cannon, (San Francisco: ASP),in press

van den Bergh, S. 2000, The Galaxies of the Local Group, (Cambridge: Cambridge Univ. Press), in press